\documentclass{elsart}

\usepackage{graphicx}
\usepackage{amssymb}
\usepackage{wasysym}

\newcommand{\be}{\begin{equation}}
\newcommand{\ee}{\end{equation}}
\newcommand{\bea}{\begin{eqnarray}}
\newcommand{\eea}{\end{eqnarray}}

\newcommand{\les}{\ell_{\hbox{\tiny ES}}}
\newcommand{\jes}{j_{\hbox{\tiny ES}}}
\newcommand{\ld}{\ell_{\hbox{\tiny D}}}
\newcommand{\tsec}{t_{\hbox{\tiny 2nd}}}
\newcommand{\tadv}{t_{\hbox{\tiny adv}}}

\newcommand{\kmax}{k_{max}}

\begin{document}

\begin{frontmatter}



\title{Nucleation and step-edge barriers always destabilize
step-flow growth of a vicinal surface}


\author{Daniele Vilone$^a$, Claudio Castellano$^a$, 
Paolo Politi$^{b,c~*}$}

\corauth{Corresponding author. {\it E-mail address:}
Paolo.Politi@isc.cnr.it}

\address{$^a$ Dipartimento di Fisica, Universit\`a di
Roma ``La Sapienza'' and Center for Statistical Mechanics and Complexity,
INFM Unit\`a Roma 1, P.le A. Moro 2, 00185 Roma, Italy}

\address{$^b$ Istituto dei Sistemi Complessi, Consiglio Nazionale delle
Ricerche, Via Madonna del Piano, 50019 Sesto Fiorentino, Italy}

\address{$^c$ Istituto Nazionale per la Fisica della Materia, UdR Firenze,
Via G.~Sansone 1, 50019 Sesto Fiorentino, Italy}

\begin{abstract}
We consider the effect of nucleation on a one-dimensional
stepped surface, finding that step-flow growth is
metastable for any strength of the additional step-edge barrier.
The surface is made unstable by the formation of a critical
nucleus, whose lateral size is related to the destabilization
process on a high-symmetry surface.
Arguments based on a critical nucleus of height two, which suggest
the existence of a fully stable regime
for small barrier, fail to describe this phenomenon.

\end{abstract}

\begin{keyword}
Crystal growth \sep
Nucleation \sep
Metastability

\PACS 81.10.Aj \sep 68.55.Ac \sep 05.70.Ln 
\end{keyword}
\end{frontmatter}


\section{Introduction}

Stability, instability and metastability are well known and widespread 
concepts in statistical mechanics, whose precise meaning may depend on the
field of application. In this article we will refer to a 
crystal surface growing by Molecular Beam Epitaxy, which has two main
growth modes:
layer-by-layer growth for a singular, high symmetry surface,
and step-flow for a vicinal, stepped surface.

Layer-by-layer growth~\cite{libroMK} 
proceeds through diffusion of newly deposited 
atoms, the nucleation of stable islands and their increase in size
via aggregation; coalescence of islands and the descent of atoms
from upper terraces lead to layer completion.
Conversely, step-flow growth~\cite{reviewJW} 
proceeds through attachment of deposited atoms to 
pre-existing steps. Stability means that steps
keep straight (absence of meandering) and of uniform density
(absence of step-bunching).
The distinction between the two growth regimes is not sharp:
in qualitative terms, step-flow dominates when
the average distance between islands, the so called diffusion length
$\ld$, is larger than the separation $\ell$ between pre-existing steps.
Therefore, if $m=1/\ell$ is the slope of the surface, 
layer-by-layer and step-flow growth occur for $m$ smaller and larger than
$1/\ld$, respectively.

The presence of an additional step-edge barrier, called
Ehrlich-Schwoebel (ES) barrier, induces a slope dependent
nonequilibrium current, $\jes(m)$, since adatoms deposited on a
vicinal terrace are more likely to be incorporated at the uphill
step~\cite{JV}.
This barrier affects the stability
properties of the surface and the current allows to separate in a more
precise way the two growth regimes:
from a continuum analysis~\cite{my_review} it turns out that
a flat surface is linearly unstable at small slope ($m<m_0$) and stable
at large slope ($m>m_0$), where $m_0$ is the slope for which
$\jes(m)$ is maximal.
Step-flow of a vicinal surface ($m>m_0$) is then linearly stable,
but this does not preclude a noise-induced destabilization
analogous to the decay of metastable states in thermodynamics.
In the present article we face this question:
How and when a stepped surface is destabilized by the nucleation of
mounds on vicinal terraces?

The question, limited to 
the case of an infinite step-edge barrier, was faced by 
Krug and Schimschak, with Kinetic Monte Carlo (KMC) simulations~\cite{Krug95}.
In the limit of infinite barrier, the vicinal surface appears
to be destabilized by the formation of islands on the terraces.
This is reasonable, in particular if thermal detachment from
steps is suppressed, but what about destabilization if we
decrease the barrier and change the slope?

An answer to such a question can be provided by an argument~\cite{KK},
based on the assumption that a mound
of height two (i.e. a mound formed by a second-layer nucleation on an island)
always leads to the destabilization of the surface.
In order to understand whether the surface gets destabilized or not,
one compares the typical time $\tsec$ required for second-layer nucleation,
with the
time needed for the first-layer island  to be reabsorbed by the
advancing step of the vicinal surface, $\tadv$.
Step-flow is assumed to be destabilized only if $\tsec < \tadv$.
Explicit evaluation of these quantities indicates that the time
$\tsec$ increases as the ES barrier goes to zero, while $\tadv$
is reduced.
It turns out then~\cite{Unpub} that the criterion implies the existence
of a minimal barrier for destabilization and, as a
consequence, that step-flow growth should be {\em fully stable} at
small barriers.%
\footnote{If we introduce the ES length $\les=\exp(\Delta E/k_BT)-1$,
where $\Delta E$ is the additional step-edge barrier and $T$ is the
temperature, the minimal barrier corresponds to the critical length
$\les^c\simeq c_0\,\ld^4/\ell^3$, with a prefactor $c_0\apprge 10$.
Taking $\ld\simeq 40$ (valid for our simulations,
see below), we get, for $m=0.1$, $\les^c \apprge 10^4$,
while we numerically see metastability down to $\les=12$.}

In this Letter we show that the above picture is not correct.
First, we present numerical evidence that the time after which the system
becomes unstable is finite for all finite values of $\les$, and it
diverges (exponentially) only for $\les \to 0$.
Then, we provide and validate a heuristic argument that elucidates 
the destabilization mechanism.

\section{The destabilization process}

We perform
Kinetic Monte Carlo simulations in one dimension of the simplest model
taking into account deposition, diffusion and
the presence of an interlayer ES barrier.
The surface is represented by a set of integer height variables $h_i$
on a lattice of $L$ sites.
An average tilt $m$ is imposed via helical boundary conditions.
The initial condition is a regular train of steps, separated by a terrace
size $\ell=1/m$.
Deposition events occur at rate $F$ on randomly selected sites.
Adatoms attempt diffusion hops to nearest neighbor sites at
rate $D$ if the neighbor belongs to the same terrace,
and at rate $D'<D$ if the atom must descend a step.
Dimers and larger islands are immobile,
and no thermal detachment of atoms from steps is allowed.
We take $F=1$ and $D=5 \times 10^5$, so that the diffusion
length is $\ld \approx 40$~(the same value used in~\cite{Krug95}).
The ES length can be written as $\les=D/D'-1$~\cite{my_review}.

\begin{figure}
\includegraphics[angle=0,width=13cm,clip]{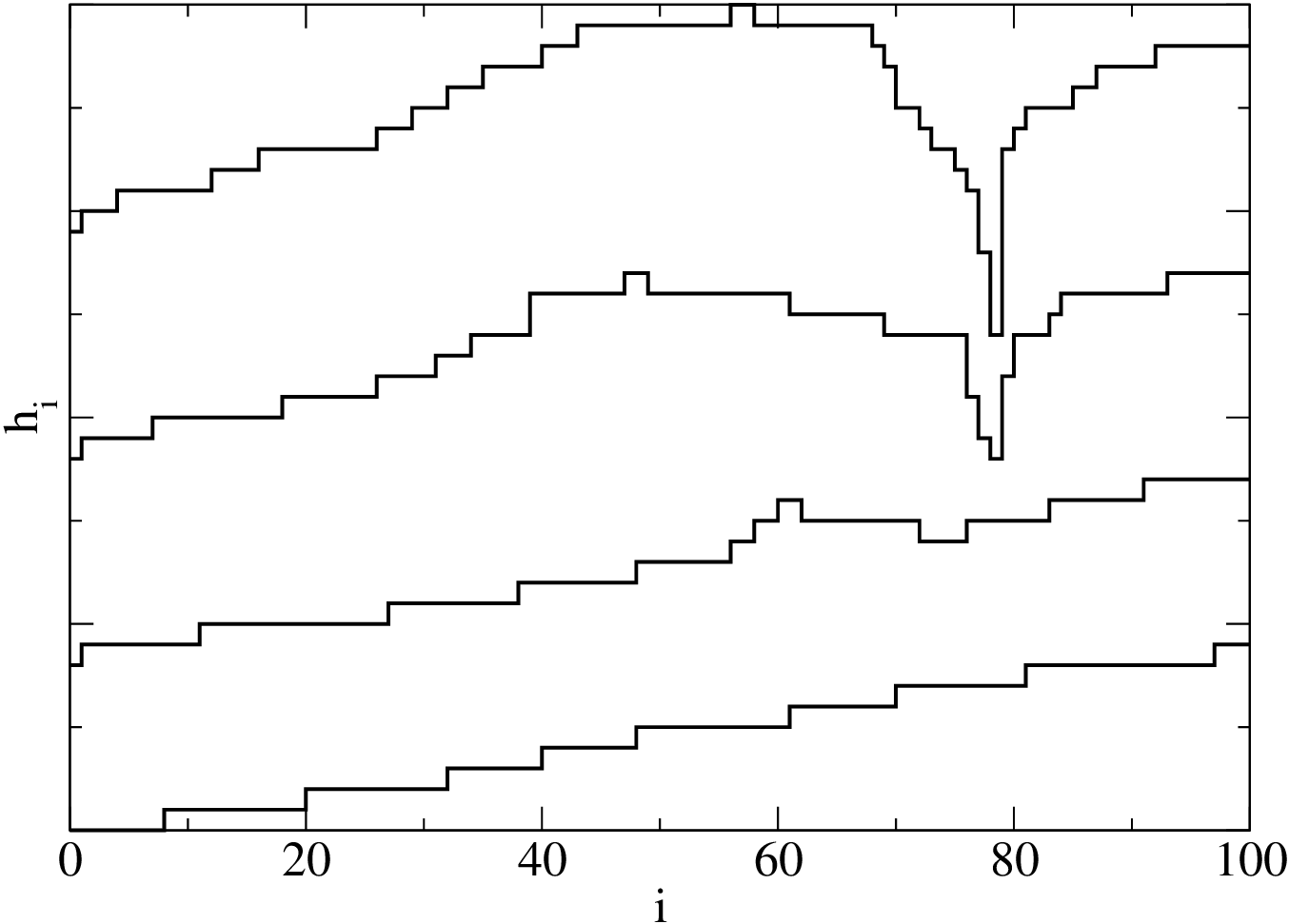}
\caption{
Schematic plot of the growing surface at different times, illustrating a
typical profile during the intermediate metastable regime (bottom),
followed by the nucleation of a dimer and the formation of a supercritical
mound.}
\label{sketch}
\end{figure}

In order to characterize quantitatively the destabilization process
we focus on the formation of the first ``critical nucleus''.
This concept arises in the continuum theory of metastability as a
localized stationary solution of the dynamics, separating stable from
unstable solutions.
In order to explain how we have identified it numerically,
let us briefly sketch how destabilization comes about (Fig.~\ref{sketch}).

During the early stages of growth the surface
is made of a train of vicinal terraces with lengths fluctuating around
the average value $\ell=1/m$. Once in a while new islands are nucleated
on them.
Some of them may grow into small bumps that are later reabsorbed by
advancing steps.
Destabilization occurs when a bump reaches some critical size
(critical nucleus) and it starts to grow irreversibly.
In principle the destabilization time $\tau$ is the moment when
mound growth becomes irreversible.
However, in a simulation such a moment is not well defined,
because mound growth is a stochastic process and even mounds which
are not reabsorbed may temporarily shrink to some extent
during their growth,
making the numerical distinction between stable and unstable mounds 
very hard.

We decided to set a height threshold at 20 layers:
when a mound reaches such a height we consider
step-flow to be destabilized and measure the destabilization time $\tau$.
There are two possible systematic errors in this way of proceeding:
in the first place, one may take as unstable a mound that is going to be
reabsorbed; secondly, one 
overestimates $\tau$, since the true destabilization of a mound typically
happens before it gets 20 monolayers high.
We have carefully taken into account these possible problems.
For all values of $\les$ and $m$ considered, we have
never seen mounds higher than 14 layers being reabsorbed.
It is therefore very likely that all mounds we have taken as unstable
(higher than 20 layers) were actually so.
For what concerns the determination of $\tau$, the time needed for a
mound to grow from height 1 to 20 is of the order of few tens of monolayers.
This may introduce a significant systematic error and invalidate the
results when $\tau$ is small, but it is irrelevant
when destabilization occurs after thousands of monolayers or more, which
is the typical case here.

The destabilization process can be split into: 
(1) the formation of a dimer on a vicinal terrace
(which occurs at a rate $1/\tau_{nucl}(L)$);
(2) the evolution of the dimer into an unstable mound (occurring
with probability $p_u$).
The total rate for the destabilization is the product of these two
quantities, so that for a system of size $L$ the mean destabilization time
$\tau(L)$ is
\be
\tau(L) = \tau_{nucl}(L) {1 \over p_u}.
\ee
$\tau_{nucl}(L)$
goes to a finite value for $\les \to 0$ and contains the whole
$1/L$ dependence on the system size. $p_u$ instead, does not depend
on $L$ and contains the relevant dependence on $\les$.

\begin{figure}
\includegraphics[angle=0,width=13cm,clip]{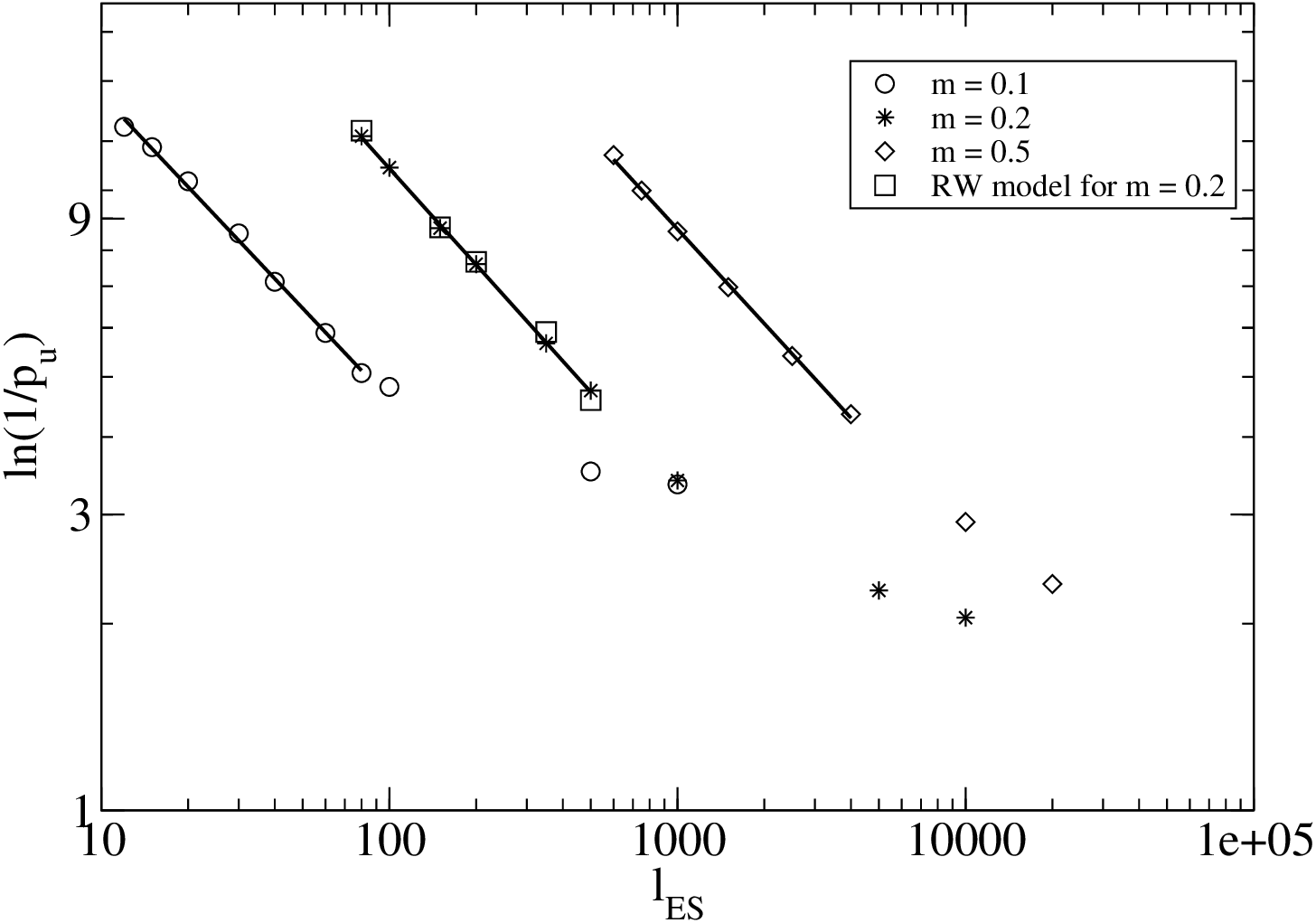}
\caption{
Double logarithmic plot of the logarithm of
$1/p_u=\tau(L)/\tau_{nucl}(L)$ as a
function of the ES length for three values of the slope $m$, computed
for a system of size $L=1000$ and for the RW model for $m=0.2$.
Time is measured in monolayers, i.e. $1/F$ is the time unit.
Statistical error bars are of the order of the symbol size.
Least-square fits (shown as lines) to the functional form
$\exp(a/\les^\gamma)$ yield
$\gamma=0.49,0.51$ and $0.51$ for $m=0.1,0.2$ and $0.5$, respectively.
}
\label{logtau}
\end{figure}

In Fig.~\ref{logtau} we plot, in a double logarithmic scale, the logarithm
of $1/p_u$ as a function of the
Ehrlich-Schwoebel length $\les$ for several values of the tilt $m$.
It is clear that the instability time diverges for $\les \to 0$
as $\exp(a/\les^\gamma)$;
there is no indication of a divergence at finite $\les$, i.e. no
indication of a fully stable region.
The value of $\gamma$ is of the order of 0.5 for all slopes considered.
For large $\les$ the curves approach a constant value
increasing with $m$.

\section{The critical nucleus}

\begin{figure}
\includegraphics[angle=0,width=13cm,clip]{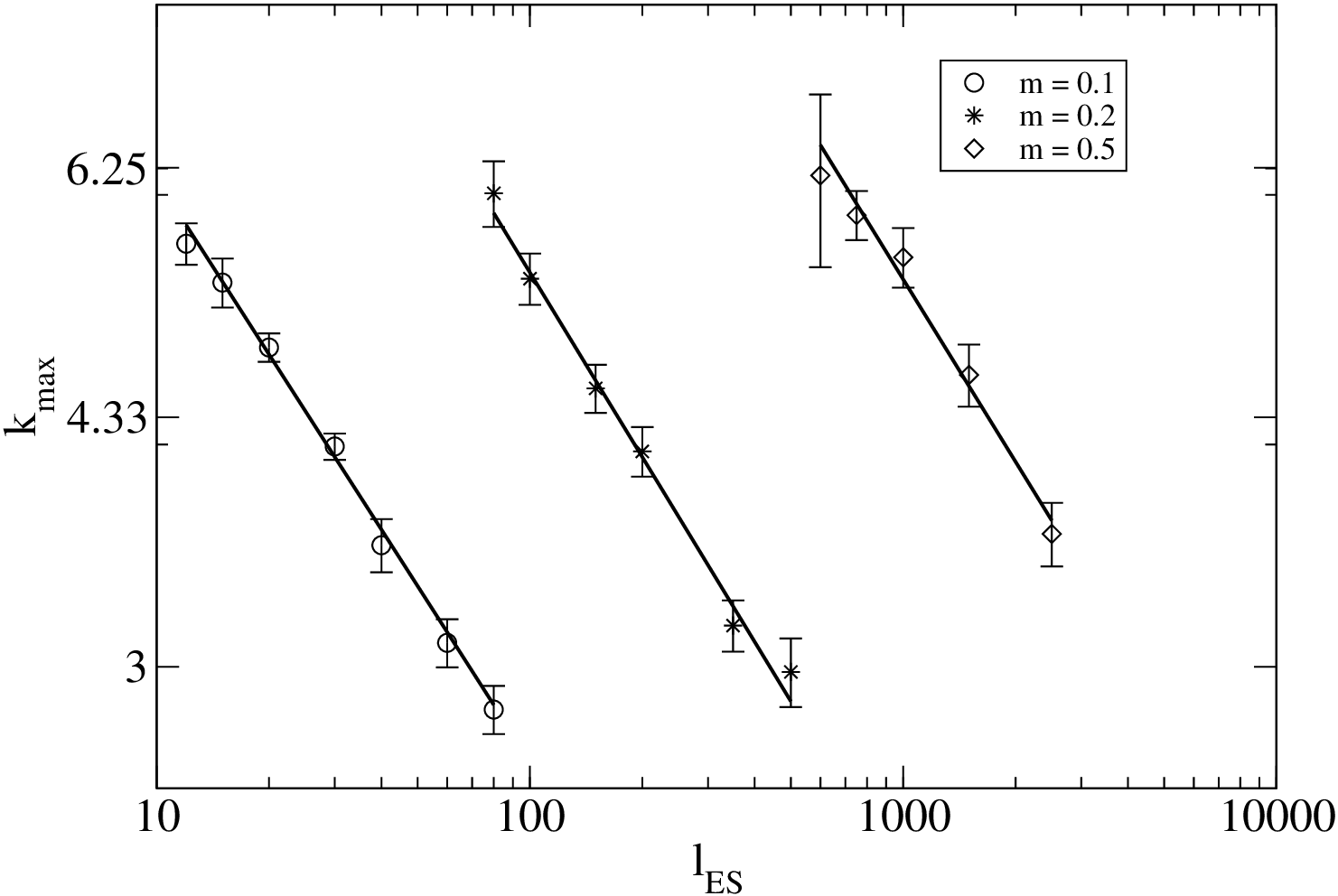}
\caption{
Double logarithmic plot of $\kmax$ as a
function of the ES length for three values of the slope $m$.
Least-square fits (shown as lines) to the functional form $\les^{-\beta}$ yield
$\beta=0.37 \pm 0.01$ for $m=0.1$,
$\beta=0.39 \pm 0.02$ for $m=0.2$ and $\beta=0.39 \pm 0.03$ for $m=0.5$.}
\label{kmax}
\end{figure}

In order to look for a theoretical explanation of the behavior of $1/p_u$
and understand
why the argument based on a critical nucleus of height two~\cite{KK} fails,
it is important to get information on the shape and the size of the
critical nucleus.
For this purpose, we compute for each realization $i$
of the dynamics, the height $k_{max,i}$ of the highest mound that is
eventually reabsorbed without diverging. After averaging over many
realizations we obtain the values $\kmax$ that are reported in Fig.~\ref{kmax}.
In the hypothesis that a critical height $k_c$ can be defined,
$\kmax$ is clearly a lower bound for $k_c$.
Its growth with decreasing $\les$ invalidates
the assumption $k_c=2$ upon which the argument of Ref.~\cite{KK} is built:
Figure~\ref{kmax} shows that, in order to become unstable,
mounds must grow higher and higher as $\les$ is reduced.

We now relate these observations to the exponential
divergence of the instability time, via a reformulation of 
the mound dynamics as a one-dimensional random walk (RW).
The formation of a mound begins with the nucleation of a dimer
on a vicinal terrace of size $\ell$ occurring at a rate $1/\tau_{nucl}(L)$.
After a dimer is formed, two possible events may occur: either the dimer
is reabsorbed, with probability $1-p_+$, or it grows higher, i.e. 
a second-layer nucleation occurs with probability $p_+$
before the advancing step catches the dimer.
In the latter case a mound of height 2 is formed and the ensuing
evolution may be the decay into a mound one-layer high or the upgrade
to a mound of height 3.
Hence the height $k$ of a mound performs a one-dimensional random walk.
Particles are introduced at $k=1$ via dimer nucleation on a terrace
and may disappear at a sink in $k=0$.
The boundary condition for high $k$ encodes the way a mound
becomes unstable.
It is possible to extract from the KMC simulations the detailed
form of $p_+(k)$~\cite{Unpub} and it is found that $p_+(k)$ grows with
$k$ from 0 to a constant value slightly larger than 1/2 for $k \ge k_s$,
with $k_s$ growing as $\les \to 0$.
Hence a critical height ($k_s$) can be defined only in a probabilistic
sense: the mound becomes unstable because once it reaches, through a rare
fluctuation, the height $k_s$ it is on average more likely for it to grow
higher than to shrink.
Using the precise form of $p_+(k)$ it is possible to determine
numerically $1/p_u$ within the RW model. The result, presented in
Fig.~\ref{logtau} shows an excellent agreement, indicating that the
RW reformulation accurately describes the mound evolution.

As just discussed, it is not fully appropriate to consider a
deterministic critical height $k_c$. However, if we approximate the true form
of $p_+(k)$ with a value $p_+^0<1/2$ for $k<k_c$ and with $1$ for $k > k_c$
the probability $p_u$ can be easily determined~\cite{Rednerbook}
\be
1/p_u = {
1-\left({1-p_+^0 \over p_+^0} \right)^{k_c}
\over
1 - {1-p_+^0 \over p_+^0} 
}
\label{tins}
\ee

Since $p_+^0$ is smaller than $1/2$ and $k_{max}$ (related to the
critical height by $\kmax=k_c-1$) is seen to increase in the
limit $\les \to 0$ (Fig.~\ref{kmax}), then
\be
1/p_u \sim (1/p_+^0-1)^{k_c} \sim \exp(a' \kmax),
\label{tau_ins}
\ee
with $a'$ depending on $p_+^0$, i.e. on the slope $m$.
Despite the approximations used in its derivation,
formula~(\ref{tau_ins}) gives a divergence $\exp(1/\les^{\beta})$
of the instability time, with $\beta$ of the order of 0.4
(see Fig.~\ref{kmax}), in rather good agreement with the KMC simulations.

\begin{figure}
\includegraphics[angle=0,width=13cm,clip]{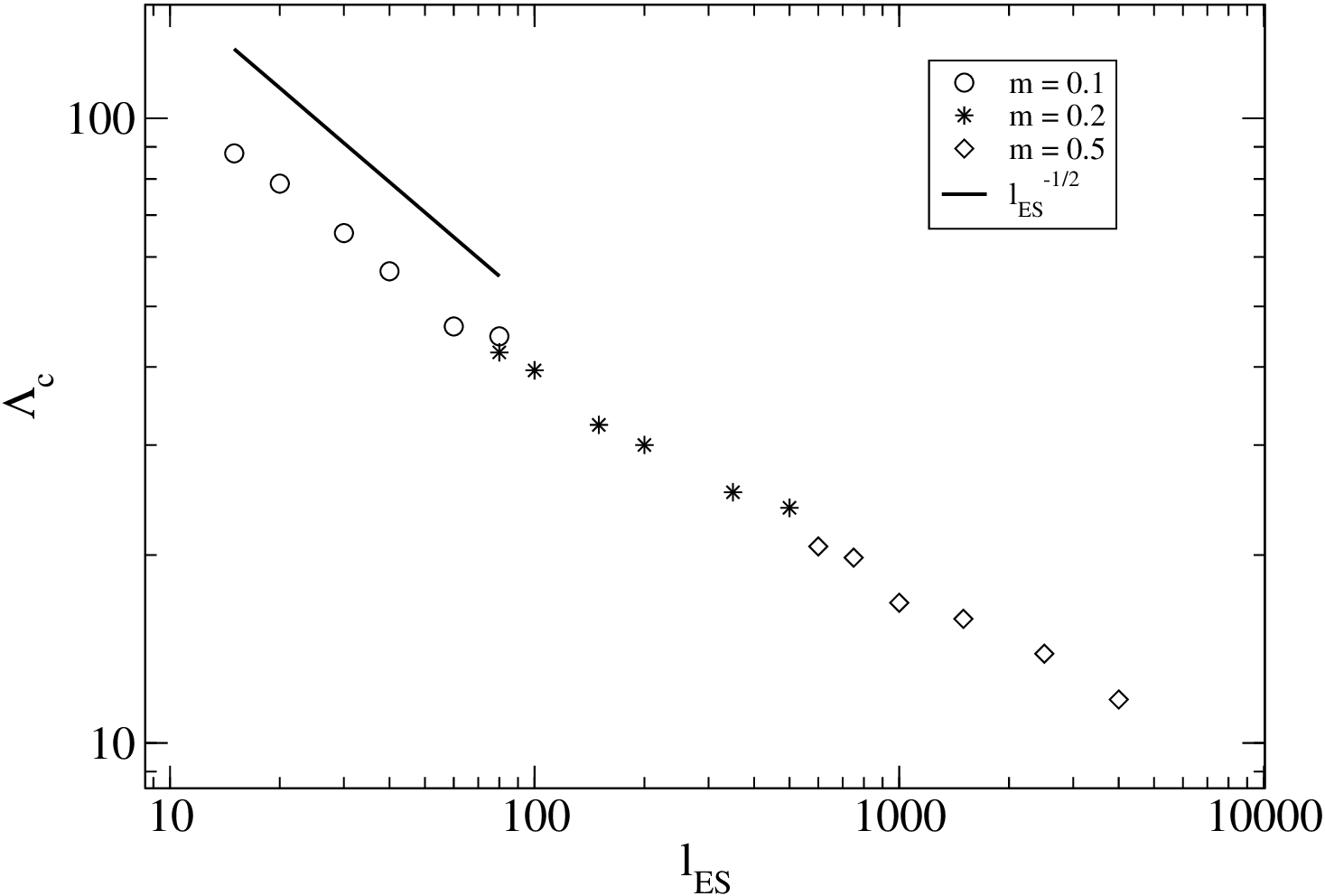}
\caption{
Double-logarithmic plot of the critical mound width $\Lambda_c$ as a
function of the ES length for three values of the slope $m$.
Statistical error bars on $\Lambda_c$ are of the order of the
symbol size.
As a guide for the eye, a divergence $\les^{-1/2}$ is also plotted.}
\label{Lambda}
\end{figure}

In order to get deeper insight into the destabilization
mechanism, let us consider the width $\Lambda_c$ of the critical nucleus.
To measure it in our simulations an unambiguous operational definition
of the width $\Lambda$ of a mound is needed.
The simplest would be to take $\Lambda$ as the base of the mound,
but nucleation events occurring along the sides make difficult a precise
identification of the base. As a consequence of that,
reasonable definitions of the base lead to strongly fluctuating quantities.
We have found that a convenient way to proceed is to take $\Lambda$ as
the size of the sub-top terrace, i.e. the second highest terrace in the mound.
We have considered critical mounds\footnote{We have considered 
the mounds that reached the threshold height 20 and taken as
$\Lambda_c$ the size of their sub-top terrace when their height was 10.}
and determined the corresponding critical width $\Lambda_c$,
as a function of $\les$ for three different values of $m$
(Fig.~\ref{Lambda}).

The most important feature of Fig.~\ref{Lambda} is that there is a unique
curve $\Lambda_c(\les)$ for all values of $m$: 
when the mound becomes unstable, the width of the sub-top terrace
depends only on $\les$ and not on the average surface tilt.
Moreover, we find that $\Lambda_c(\les)$ grows with decreasing $\les$ as a
power-law with exponent not far from 1/2.

This phenomenology can be consistently interpreted by assuming that
a mound formed on a vicinal surface is similar to a mound appearing
on a singular surface of size $L$ of the order of $\Lambda$.
Let us recall that for a high symmetry
surface there exists a critical size $L_c(\les)$ such that a surface
larger than $L_c$ is linearly unstable, while, if the opposite is true,
the instability is hindered and the planar surface is stable~\cite{reviewJK}.
For mounds on stepped surfaces this threshold is reflected in the
existence of a critical width $\Lambda_c(\les)$, close to,
although not necessarily coinciding with $L_c(\les)$.
As long as a mound on a stepped surface is small it tends on average to be
reabsorbed, exactly as perturbations on a high-symmetry surface with $L<L_c$
tend to disappear.
However, if nucleation events occur more rapidly than on average, so that
its lateral size becomes larger than
\be
\Lambda_c(\les) \approx L_c(\les),
\label{criterion}
\ee
then the mound becomes unstable.

This scenario is fully consistent with the numerical results presented
in Fig.~\ref{Lambda}: in particular, the independence of
$\Lambda_c(\les)$ on $m$ and the
divergence of $\Lambda_c(\les)$ for $\les \to 0$
(consistent with the fact that, in a continuum treatment,
$L_c \approx 1/\sqrt{\les}$) strongly support the validity of
Eq.~(\ref{criterion}).

The connection between the instability on a high-symmetry surface 
(which occurs independently of the amplitude of fluctuations) and
on a stepped surface has an additional implication: the relevant
quantity is the width and not the height of the mound.
The observed increase of the critical nucleus height as $\les \to 0$,
is a consequence of the more fundamental growth of its lateral size.
We have numerically tested this by taking mounds larger than
$\Lambda_c(\les)$ and drastically reducing their height to 2 or 3 layers.
This modification does not change their unstable nature.
This observation confirms that the height of a mound is not the relevant
variable for the instability. Nevertheless, the width
and the height of a mound are dynamically correlated and for this reason 
studying the height of a mound is useful even if its stable or unstable
character depends on its width. 

\section{Conclusions and open questions}

In summary, we have shown that any additional step-edge barrier makes
step-flow growth unstable with respect to the formation of mounds.
The time needed for this instability to take place diverges exponentially
as $\les$ goes to zero. This is a consequence of the divergence
of the size of the critical nucleus needed to destroy the metastable step-flow.
This divergence is, in its turn, related to the process of destabilization
of a high-symmetry surface.

There are several open, or not fully understood, questions which
we are currently investigating~\cite{Unpub}.
The first issue is to consider how our arguments can be 
extended to two dimensions. When steps are lines rather than points,
the additional possibility of meandering arises and, as shown by
Kallunki and Krug~\cite{KK}, this is a second mechanism for
destabilizing step-flow, which is effective even if nucleation
is completely suppressed. As for the destabilization induced by
nucleation, the $k_c=2$ approximation still
provides~\cite{KK} a threshold $\les^c$.
Since this result is based on a constant critical height $k_c$,
independent from $\les$, it is most likely that the approximation
fails in two dimensions as well. 

With regard to numerics, we want to analyze the
crossover between the linear unstable regime ($m<m_0$) and the
metastable regime ($m>m_0$).
This crossover is made subtle by the $\les-$dependence of the slope $m_0$.
In addition, we want to study the effect of thermal
detachment from steps. Does it have purely quantitative effects
or does it change in a profound way the destabilization picture? Finally,
from the analytical point of view, we have found that standard continuum
(Cahn-Hilliard type) theories are not suitable for describing
the $\les$ dependence of
the destabilization process of the vicinal surface, not even
in a qualitative way~\cite{Unpub}.
The possibility to give an alternative continuum
description is an important open issue.



\end{document}